# Nonlinear optical detection of the orbital angular momentum of light

Ju-Young Kim

Center for Molecular Spectroscopy and Dynamics, Institute for Basic Science (IBS), Seoul 02841, Republic of Korea

Minhaeng Cho*

Center for Molecular Spectroscopy and Dynamics, Institute for Basic Science (IBS), Seoul 02841, Republic of Korea

Department of Chemistry, Korea University, Seoul 02841, Republic of Korea

* mcho@korea.ac.kr

## Abstract

Optical beams carrying orbital angular momentum (OAM) have gained significant interest due to their unique properties, enhancing various communication systems and enabling applications such as the characterization of material or molecular chirality. Generating and detecting the OAM of light is thus crucial for numerous applications but poses challenges. This paper proposes a method utilizing stimulated Raman scattering to detect the magnitude of OAM. By exploring the strong Raman coupling between the pump and Stokes beams in higher-order Laguerre-Gauss modes, we demonstrate the discrimination of different optical vortex beams through nonlinear optical measurements. Numerical and experimental results support the feasibility of this approach, potentially advancing optical communication and sensing technologies.

## 1. Introduction

The study of optical beams carrying orbital angular momentum (OAM), also known as optical vortices (OVs), has attracted significant attention among researchers due to their unique properties[1]. These beams are characterized by a helical wavefront and a phase proportional to exp(-$il\varphi$), where $l$ is the integer representing the topological charge (TC). This versatile nature allows optical OAM modes to serve as an orthogonal basis, enhancing various communication systems[2], including fiber optic[3], free-space[4, 5], and quantum communications[2, 6]. Moreover, the phase of optical vortices plays a crucial role in analyzing laser field polarizing properties, enabling laser beam detection and characterization[7], and characterizing material chirality[8].

The applications of optical vortices span a wide range of fields[9-11], from optical tweezers and spanners to sensing technologies and object motion detection[12-15]. Consequently, there is an increasing demand for efficient methods to generate these beams. Various techniques have been developed and categorized into spatial and fiber-based methods. Diffractive optical components, spiral phase plates, spatial light modulators, metasurfaces, and photonic integrated circuits are among the devices utilized for helical beam generation, each offering distinct advantages in terms of efficiency and scalability[10, 11].

The detection of OAM of light is the reverse process of the OAM-light generation. It is important for various applications in fields like optical communication and transmission because optical vortex beams (OVBs) carrying OAM offer a higher degree of freedom, enhancing the capabilities of communication systems. Free-space optical communication, in particular, has gained attention due to its flexibility, security, and large signal bandwidth[4, 5, 16]. However, detecting OAM at the receiver side still poses challenges.

Numerous research efforts have been devoted to developing techniques for detecting the OAM of OVBs [17-20]. While these advancements have significantly contributed to the field, challenges remain, including computational complexity, simulation requirements, and the need for beam restoration in disturbed transmission environments[21, 22]. Thus, further research is needed to develop more robust and efficient detection methods to improve detection fidelity.

In this paper, we propose a method utilizing a nonlinear optical process to detect the magnitude of OAM. Among different nonlinear optical phenomena, we consider the stimulated Raman scattering (SRS) process as a means to extract information on the OAM of light. While the exploration of SRS with high-intensity light has been thorough both theoretically and experimentally[23, 24], it has garnered significant attention for its applications in bioimaging over the past two decades[25]. Unlike fluorescence-based microscopy techniques, SRS or coherent anti-Stokes Raman scattering (CARS) microscopy allows for obtaining spatial distribution and functional information about the primary chemical components of biological systems in a label-free manner[26, 27].

Recently, solving the Raman-coupled wave equations for the pump and Stokes beams, we demonstrated that the pump beam in a $TEM_{00}$ mode can be focused when its conjugate Stokes beam is an annular beam with a non-zero TC [28]. Such Raman coupling between the two beams with different azimuthal indices occurs only when they overlap in space. As the TC of a beam with a fixed radial index increases, the radius of the doughnut ring increases. Consequently, one can discriminate different OVBs by allowing them to be Raman-coupled to a reference beam of which intensity distribution is pre-determined, e.g., $TEM_{00}$ Gaussian beam. This approach, exploiting a nonlinear optical measurement for OAM detection, has critical advantages compared to the conventional method utilizing phase plates, spatial light modulators, or specially designed nanostructures. First, the method is an all-optical technique without relying on the use of amplitude-phase-modulating materials because the nonlinear optical property (four-wave-mixing) of Raman-active material is used to couple the vortex beam (with unknown TC) and reference (with known TC) beam and only the reference beam is under measurement. Second, the approach is a non-destructive method since the photons in the vortex beam are not annihilated by photo-detectors or photocurrent measurement devices, which makes the approach unique. Third, due to the increase in vortex Stokes beam intensity (while its TC preserves) due to the stimulated Raman gain process, it can be used for amplifying the vortex beam carrying optical information without alteration of its TC.

Despite its uniqueness and advantages, this approach, nonlinear optical coherent Raman detection of the OAM of light, has not been studied nor alluded to before. In this paper, we present numerical results, solving the Raman-coupled wave equations of the Laguerre-Gaussian (LG) pump and Stokes beams, where the latter corresponds to the OVB used in optical communication. Experimental results further confirm that the stimulated Raman loss of the pump increases as the spatial overlap of the Gaussian pump and vortex (Laguerre-Gauss) Stokes beams increases, demonstrating the practical possibility of OVB detection by nonlinear optics. We anticipate that the present work could contribute to the ongoing efforts in advancing optical communication and sensing technologies.

## 2. Theoretical description

### A. Coupled wave equations for the pump and Stokes beams

We consider the following wave equation for $\vec{E}$ with nonlinear polarization $\vec{P}^{(3)}$, a source term:[23, 24, 29-31]

$$\nabla^2 \vec{E} - \frac{n^2}{c^2}\frac{\partial^2}{\partial t^2}\vec{E} = \frac{4\pi}{c^2}\frac{\partial^2}{\partial t^2}\vec{P}^{(3)}, \tag{1}$$

where $n$ and $c$ represent the refractive index and the speed of light, respectively. This work focuses on the SRS process within the paraxial approximation. The electric fields propagating along the $z$ direction can be expressed as follows, for $j$ representing the pump and Stokes fields:

$$E_j(x,y,z) = \mathcal{E}_j(x,y,z)e^{ik_j z - i\omega_j t}. \tag{2}$$

The challenge lies in determining the spatial amplitudes $\mathcal{E}_j(x,y,z)$ and intensity distributions $\frac{1}{2}\varepsilon_0 cn|\mathcal{E}_j(x,y,z)|^2$ of the pump and Stokes beams, specifically when their input beam profiles are tailored using spiral phase plates, holographic gratings, metamaterials, or spatial light modulators[9]. Here, $\varepsilon_0$ represents the vacuum permittivity, and $\omega$ is the angular frequency of the beam. With that, we will consider a two-beam SRS process involving the pump and Stokes beams as the high- and low-frequency fields, respectively [26].

Inserting Eq. (2) into the wave equation (1) and invoking the slowly varying envelop approximation[29], which implies that $\frac{\partial^2 \mathcal{E}_j}{\partial z^2} \ll 2ik_j \left(\frac{\partial \mathcal{E}_j}{\partial z}\right)$ where $k_j$ is the wavenumber, yield

$$\nabla_t^2 \mathcal{E}_p + 2ik_p \left(\frac{\partial \mathcal{E}_p}{\partial z}\right) = -\frac{4\pi k_p^2}{n_p^2} \chi_R^{(3)^*} |\mathcal{E}_S|^2 \mathcal{E}_p \tag{3}$$

$$\nabla_t^2 \mathcal{E}_S + 2ik_S \left(\frac{\partial \mathcal{E}_S}{\partial z}\right) = -\frac{4\pi k_S^2}{n_S^2} \chi_R^{(3)} \left|\mathcal{E}_p\right|^2 \mathcal{E}_S \tag{4}$$

where $\nabla_t^2 \equiv \frac{\partial^2}{\partial x^2} + \frac{\partial^2}{\partial y^2}$.

It is worth noting that the coefficient on the right-hand side of the wave equation for the pump (Stokes) beam includes $\boldsymbol{\chi}_{\boldsymbol{R}}^{(\mathbf{3})^*}$ ($\boldsymbol{\chi}_{\boldsymbol{R}}^{(\mathbf{3})}$), where $\boldsymbol{\chi}_{\boldsymbol{R}}^{(\mathbf{3})}$ corresponds Raman susceptibility[6]. Since the imaginary part of $\boldsymbol{\chi}_{\boldsymbol{R}}^{(\mathbf{3})}$ is negative, the intensity of the pump (Stokes) beam is decreased (increased) by the SRS process. In other words, the pump undergoes a stimulated Raman loss (SRL) through the SRS process, while stimulated Raman gain (SRG) occurs to the Stokes beam. It is evident that the SRS process couples the two beams. The amplitudes, phases, and, consequently, intensity distributions of these beams are interrelated, suggesting the potential to extract information about the beam properties of either the pump or Stokes beam by controlling or using the other.

**B. Coupled equations for the LG mode expansion**

Now, we will focus on cases where the intensity distributions of the pump and Stokes beams exhibit cylindrical symmetry. Consequently, the amplitudes of these beams can generally be expanded in a complete set of orthonormal LG modes in a cylindrical coordinate system. Note that the TCs of the incident pump and Stokes beams denoted as $l_p$ and $l_S$, respectively, are preserved and do not mix with other optical vortex modes having different TCs.

To measure the OAM of light using the SRS process, we consider the target vortex light with a TC of $l_c$ as Stokes beam ($l_S = l_c$), which is Raman-coupled with the reference pump beam. The carrier frequency ($\omega_p$) of the pump is determined by the Stokes beam frequency ($\omega_S$) and the Raman-active mode frequency ($\omega_{vib}$) of material chosen for the optical detection of OAM, i.e., $\omega_p = \omega_S + \omega_{vib}$. The initial beam profiles are assumed to be

$$\mathcal{E}_p(r,\phi,z_0) = P_0^{l_p}(z_0)U_0^{l_p}(r,\phi,z_0) \tag{5}$$

$$\mathcal{E}_S(r,\phi,z_0) = S_0^{l_c}(z_0)V_0^{l_c}(r,\phi,z_0), \tag{6}$$

where $z_0$ represents the *z*-position of the front surface of the Raman medium. $P_n^{l_p}(z)$ and $S_n^{l_S}(z)$ are the z-dependent expansion coefficients, related to Raman loss and gain, respectively. $U_n^l(r,\phi,z)$ and $V_n^l(r,\phi,z)$ represent the $LG_{nl}$ modes for the pump and Stokes beams, respectively, with *n* and *l* being the radial and rotational mode indices, respectively.

Figure 1(a) illustrates these initial beam intensity profiles for $l_p = 0$ and $l_c \neq 0$ before and after SRS, which was theoretically proposed to demonstrate the feasibility of super-resolution Raman imaging [32, 33]. And Figure 1(b) is a scheme of the SRS optical setup.

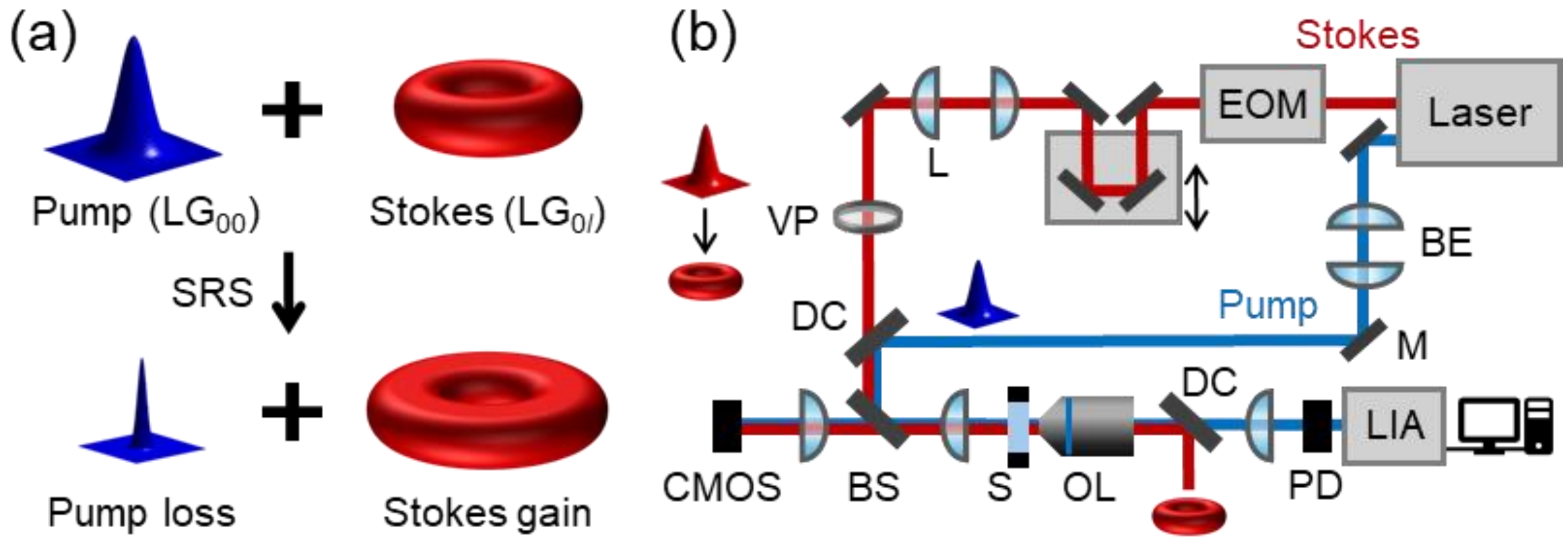

Figure 1. (a) Illustration of the SRS process with a Gaussian ($LG_{00}$) pump beam and a doughnut-shaped ($LG_{0l}$) Stokes beam. After the SRS process, the pump (Stokes) beam becomes weaker (stronger) in intensity, while topological charges are preserved. The pump beam is axially focused when the TC of the conjugate Stokes beam is larger than zero. (b) Schematic configuration of SRS setup. EOM: electro-optical modulator, BE: beam expander, L: lens, M: mirror, VP: vortex plate, DC: dichroic mirror, BS: beam splitter, S: sample stage, OL: objective lens, LIA: lock-in amplifier, PD: photodiode. While SRL of the pump is detected by a PD, the OAM of the Stokes preserves (non-destructive).

**3. Results and discussion**

To numerically solve the coupled wave equations for the pump and Stokes beams with the initial conditions provided in Eqs. (5) and (6), we used the parameters outlined in Table S1. In summary, the wavelengths of the pump and Stokes beams are 800 and 1000 nm, respectively. The Rayleigh ranges of the two beams are 3.7 and 4.62 mm, respectively. The thickness of the Raman medium is 83.2 mm, i.e., $L$=41.6 mm. The Raman gain coefficient is $2.489 \times 10^{-18}$ m/V$^2$, and the initial amplitudes of both $P_0^0(z_0)$ and $S_0^{l_c}(z_0)$ at $z_0$ $(= -L)$ are 0.752 MV in Figures 2 or 0.238 MV in Figure 3(a).

With the parameters presented above, we performed finite-difference calculations of the coupled wave equations. The lateral intensity distributions of the transmitted pump beam and OVB at $z = L$ are shown in Figure 2. Due to the significant spatial overlap between the Gaussian pump beam and $\mathrm{LG}_{0l_c}$ OVB with small TCs ($l_c$), the transmitted pump intensity experiences considerable attenuation, attributed to SRL, leading to deviation from the initial Gaussian shape in its intensity distribution. Conversely, the transmitted Stokes beam intensity increases, with its spatial distribution remaining unchanged when the TC is large (Figure 2(a)).

As demonstrated in Ref.[28], the SRS process does not induce couplings between LG modes that have different TCs but causes the mixing of LG modes with different radial indices. This preservation of OAM of light is understandable by noting that an isotropic Raman medium does not mix beams with different helicities due to its centrosymmetric and achiral properties. This conservation and non-destructiveness of the OAM of light represent a valuable advantage of the present approach in transmission geometry, in addition to amplifying the transmitted OVB carrying information via SRG without altering its original TC.

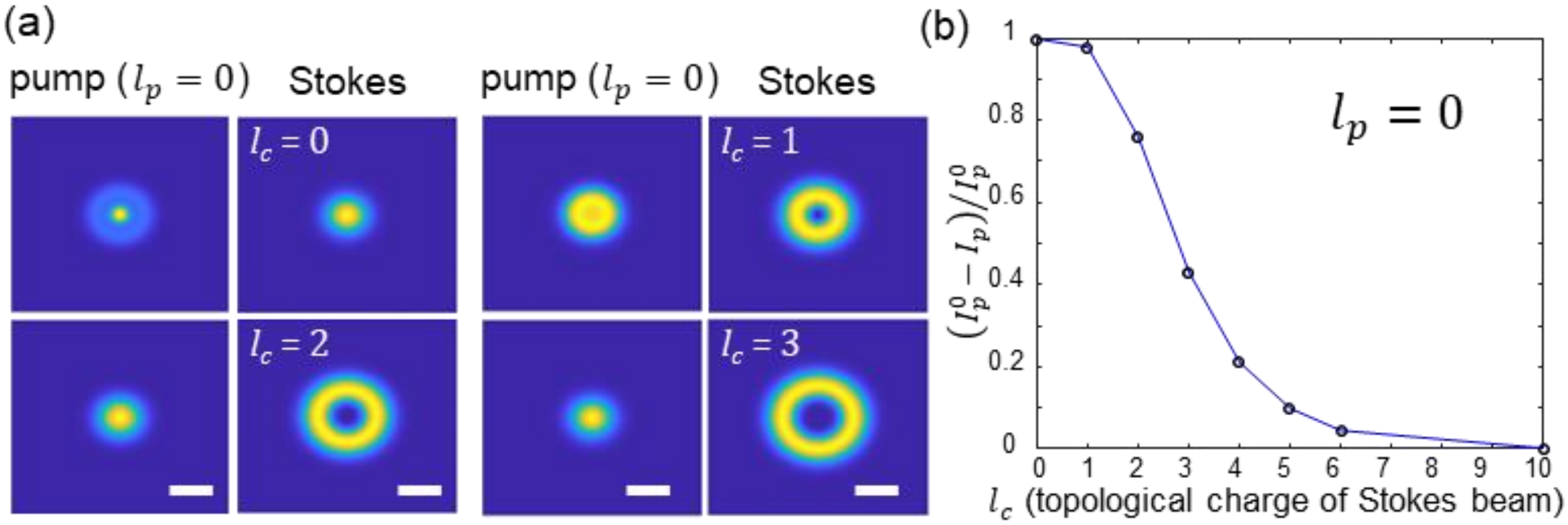

Figure 2. (a) Numerically calculated lateral intensity distributions of the transmitted pump (left column) and vortex (right column) beams at $z = L$ (= 41.6 mm). The initial pump beam at $z = -L$ is an $LG_{00}$ mode. The initial vortex beam at $z = -L$ is assumed to be an $LG_{0l_c}$ mode with $l_c$ = 0-3. The scale bar is 16.64 mm. (b) The laterally integrated intensity $I_p$ $(z = L)$ was calculated for $l_c$ = 0-10 (x-axis). Here, the ratio $(I_p^0 - I_p)/I_p^0$ versus $l_c$ is plotted.

To assess the extent of SRL of the pump beam, we calculated the integrated (over *r* and *f*) intensity of the pump beam at the end of the Raman medium at $z = L$, which is denoted as $I_p$. The integrated intensity of the reference pump beam in the absence of the Stokes beam (OVB) is denoted as $I_p^0$. The relative SRL of the pump beam intensity can be represented by the ratio $(I_p^0 - I_p)/I_p^0$, a measurable quantity using a lock-in amplifier with modulation of the OVB. If the intensities of incoming Stokes OVBs are different for varying TCs, the measured pump loss signal needs to be normalized.

Figure 2(b) shows that for OVBs with small $l_c$ values, the strong SRL of the $LG_{00}$ pump is induced by extensive SRS in the spatially broad overlapping region of the two beams, which leads to the intensity ratio being close to unity. As the TC of OVB increases, the extent of beam overlap decreases (Figure 2(a)), resulting in a weaker SRL of the pump beam. Consequently, the intensity ratio decreases and approaches zero.

To determine the unknown $l_c$ value of the OVB, one can scan the TC $l_p$ of the pump beam and measure the SRL signal, $(I_p^0 - I_p)/I_p^0$. When $l_c$ = 0, 2, 4, 6, and 8 for the OVBs carrying information, the pump beams with varying $l_p$ values undergo SRL processes differently due to the varying spatial overlaps between the two beams at the focus. The $l_p$ value at the maximum of $(I_p^0 - I_p)/I_p^0$, denoted as $l_p^*$, is linearly correlated with $l_c$ when $l_c$ is smaller than 5 (Figure 3(a)). The deviation from a linear line between $l_p^*$ and $l_c$ is because of the difference in the spatial profiles, e.g., beam waists, Rayleigh ranges, etc., of the pump and Stokes beams with different wavelengths. The correlation (Figure 3(a)) between $l_p^*$, experimentally measurable quantities, and the $l_c$ of OVB indicates that the OAM of light can be characterized using nonlinear optical measurements. Nonetheless, the dependence of SRS on the pump and Stokes beam intensities would necessitate an appropriate choice of Raman gain material, thickness, beam waist, confocal parameter of focusing device, etc.

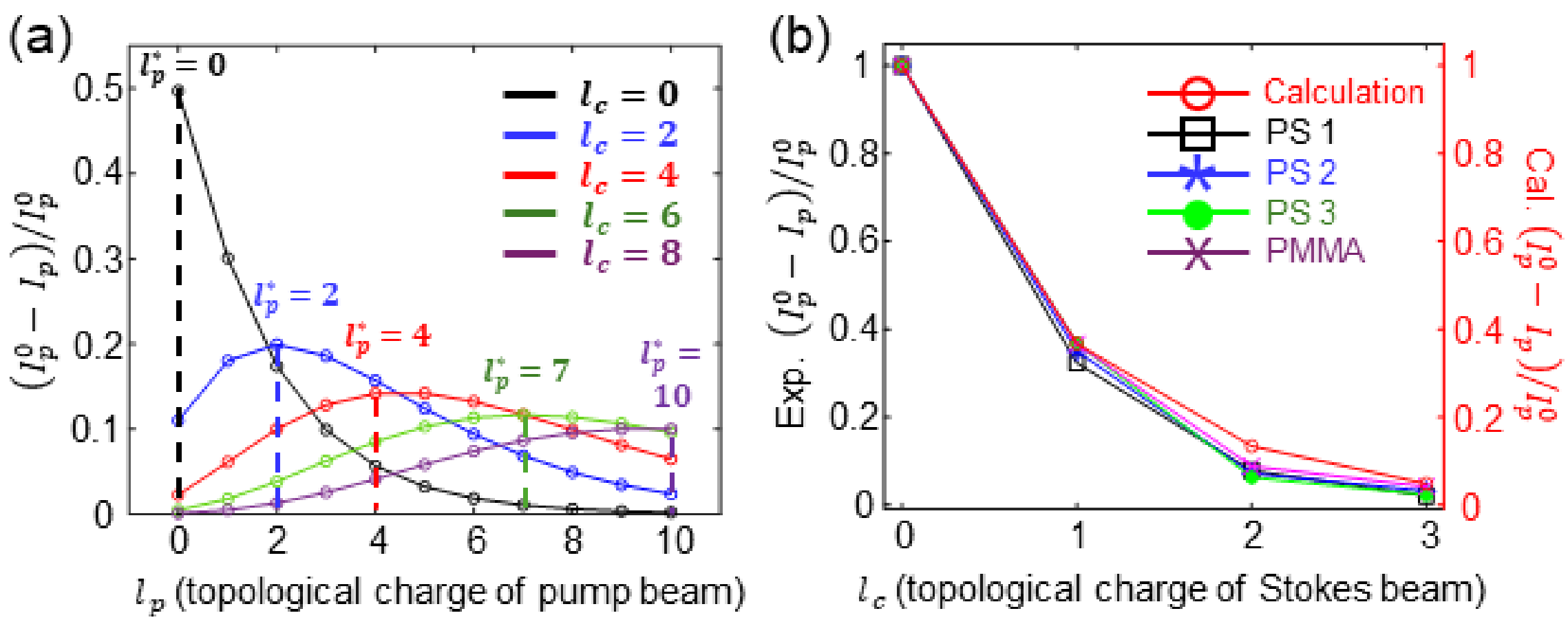

Figure 3. (a) $(I_p^0 - I_p)/I_p^0$ versus $l_p$ of the pump beam. The pump is a vortex beam with a TC of $l_p$. For the OVB, which is the Stokes beam carrying information, has a TC of $l_c$ (=0, 2, 4, 6, or 8). Then, for varying $l_p$, the SRL of the pump is measured and plotted with respect to $l_p$. The $l_p^*$ at the peak of $(I_p^0 - I_p)/I_p^0$ is identical to $l_c$ when $l_c \leq 4$. However, the $l_p^*$ is larger than $l_c$ when $l_c \geq 5$, which is due to the difference in the pump and Stokes (OVB) beam wavelengths. (b) Calculated $(I_p^0 - I_p)/I_p^0$ (red, ○) compared to the experimental values resulting from the SRS of the pump ($l_p = 0$) and Stokes ($l_c = 0, 1, 2,$ and 3) beams, where the Raman materials are PS and PMMA (magenta, X) films. PS1, PS2, and PS3 utilized varying pump intensities (black, □; blue, *; green, ●), with calculations based on PS parameters (see Table S2).

To experimentally demonstrate the changes in SRL depending on the OAMs of its conjugated Stokes beam, which is numerically calculated above, we measured $(I_p^0 - I_p)/I_p^0$ by using our home-built Raman setup (Figure 1(b)). The TCs of the Stokes beam are modulated by the vortex plates ($l_c$= 0, 1, 2, and 3). Polystyrene (PS) and poly(methyl-methacrylate) (PMMA) were chosen as Raman materials. For PS, the varying pump intensities were utilized (0.30, 0.27, and 0.21 mW/mm$^2$).

In Figure 3(b), the relative SRL of the pump beam measured after the SRS process is plotted with respect to $l_c$ of the OVB, which is the Stokes beam in this case. As predicted, the SRL decreases as the $l_c$ of the Stokes beam, or equivalently the radius of the doughnut-shaped ring, increases, proving that the spatial overlap between the Gaussian pump and the OVB reduces because the higher $l_c$ OVBs have larger inner holes (Figure 2). We compare the calculation results of SRL with the experiments. Note that, despite variations in materials (PS and PMMA) and varying pump powers (indicated as PS1, PS2, and PS3), the data consistently align with the results of the calculations. These experimental results and theory reveal that even with variations in intensity within certain limitations and the use of commonly employed polymers, the method can reliably measure the OAM of vortex light, demonstrating its potential for practical applications.

## 4. SUMMARY

In summary, we introduced a theoretical framework and presented numerical calculation results along with the experimental verification of the feasibility of detecting the OAM of vortex light through SRS measurements with vortex pump beams. It was demonstrated that the degree of SRL of the pump beam depends on the OAMs of both the pump and OVB. This sensitivity could be harnessed to ascertain the magnitude, rather than the sign, of the optical vortex beam utilized in optical communication. Given these unique features of nonlinear optical (coherent Raman) detection, we believe that this approach will find an exceptional use for the relay and amplifier in optical vortex beam communication.

**Acknowledgments.** We thank the Institute for Basic Science for financial support (IBS-R023-D1).